\documentclass{iaus}
\usepackage{graphicx}
\usepackage{natbib}
\bibpunct{(}{)}{;}{a}{}{,}
\title[Spectral Analysis of Hot Post-AGB Stars] 
{Non-LTE Spectral Analysis\\of Extremely Hot Post-AGB Stars:\\Constraints for Evolutionary Theory}

\author[Rauch et al\@.]   
{Thomas Rauch$^1$, Klaus Werner$^1$, Marc Ziegler$^1$,\\Lars Koesterke$^2$, Jeffrey W\@. Kruk$^3$}

\affiliation{$^1$Institute for Astronomy and Astrophysics,
             Kepler Center for Astro and Particle Physics,
             Eberhard Karls University, 
             Sand 1,
             72076 T\"ubingen, 
             Germany\\
             e-mail: rauch@astro.uni-tuebingen.de\\[\affilskip]
             $^2$Texas Advanced Computer Center, University of Texas, Austin, TX 78712, USA\\[\affilskip]
             $^3$Department of Physics and Astronomy, Johns Hopkins University, Baltimore, MD 21218, USA
            }

\pubyear{2008}
\volume{252}  
\pagerange{1--8}
\date{?? and in revised form ??}
\jname{The Art of Modelling Stars in the 21$^{st}$ Century}
\editors{L\@. Deng, K\@. L\@. Chan, C\@. Chiosi, eds.}
\begin{document}

\maketitle

\begin{abstract}
  Spectral analysis by means of Non-LTE model-atmosphere techniques 
has arrived at a high level of sophistication: fully line-blanketed 
model atmospheres which consider opacities of all elements from H to 
Ni allow the reliable determination of photospheric parameters of hot, 
compact stars. Such models provide a crucial test of stellar evolutionary theory: 
recent abundance determinations of trace elements like, e.g., F, Ne, Mg, 
P, S, Ar, Fe, and Ni are suited to investigate on AGB nucleosynthesis. 
E.g., the strong Fe depletion found in hydrogen-deficient post-AGB 
stars is a clear indication of an efficient s-process on the AGB where Fe 
is transformed into Ni or even heavier trans iron-group elements. 
  We present results of recent spectral analyses based on high-resolution 
UV observations of hot stars.
\end{abstract}

\firstsection 
\section{Introduction}
\label{sect:introduction}

In the last decades, our picture of post-AGB stellar evolution has been greatly improved.
The ``standard'' evolution, i.e\@. the hydrogen-rich sequence, has been understood
in the early eighties of the last century by comparison of spectral analysis and
evolutionary models. The spectral analysis of the hottest post-AGB stars with
effective tempe\-ratures ($T_\mathrm{eff} > 100,000\,\mathrm{K}$) was hampered by
the lack of appropriate model atmospheres which considered deviations from the
local thermodynamic equilibrium (LTE) playing an important role in the photospheres
of these stars. The development of such Non-LTE model atmospheres is briefly summarized in
Sect.~\ref{sect:models}. 

  Spectral analyses of hot post-AGB stars have shown then that about a quarter of these
are hydrogen-deficient \citep{werner06}. The ``born-again post-AGB star'' scenario by
\citet{iben83} is -- in general -- able to explain the evolution of these stars. A
final thermal pulse (TP, re-ignition of the helium shell) brings the star back to the
AGB and it has a second, helium-burning  post-AGB phase. However, the mechanism to dispose of 
the entire hydrogen-rich envelope was unclear. Present evolutionary calculations which 
consider the mixing and burning processes during the helium-shell flash in detail, are able to 
explain the hydrogen deficiency. 

\begin{figure}
 \includegraphics{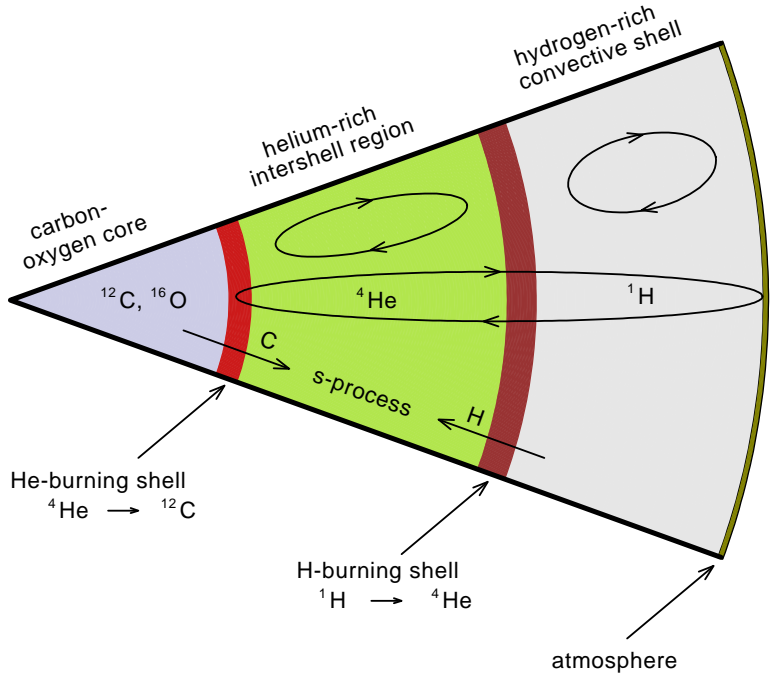}\centering
  \caption{Inner structure of an AGB star. Note that the convection zone from the surface to the bottom
           of the helium-burning shell is established during a TP only.}
  \label{fig:structure}
\end{figure}

The amount of remaining hydrogen is depending on the particular 
time when the TP occurs. Still on the AGB (AGB Final Thermal Pulse, AFTP), the masses
of the hydrogen-rich envelope and the helium-rich intershell layer (Fig.~\ref{fig:structure})
are about equal, being roughly $ 10^{-2}\,\mathrm{M_\odot}$. The TP mixing
results in $\approx 40\,\%$ hydrogen content (by mass) at the stellar surface. This is
detectable in the observations. After the departure from the AGB, the mass of the hydrogen-rich
envelope is much lower ($M_\mathrm{H} \approx 10^{-4}\,\mathrm{M_\odot}$). If the TP occurs at still constant
(high) luminosity (Late Thermal Pulse, LTP), i.e\@. nuclear burning is still ``on'', the
mixing during the flash reduces the surface hydrogen to $\approx 1\,\%$ which is not detectable at
the high surface gravity ($\log g \approx 5 - 8$ in cm/sec$^2$). A TP at already declining luminosity,
i.e\@. there is no entropy barrier due to the hydrogen-burning shell, will mix down the hydrogen-rich 
envelope to the bottom of the now re-ignited helium shell where hydrogen is burned and the star becomes
hydrogen free.

The mixing process during the TP brings intershell matter up to the stellar surface and, thus,
allows a direct view on it. Spectral analysis allows then to conclude on details of both,
mixing as well as nuclear processes like, e.g\@. the s-process, in AGB stars. This is a crucial
test for stellar evolutionary models.

\section{Model atmospheres and atomic data}
\label{sect:models}

The spectral analysis of hot stars requires suitable Non-LTE model atmospheres\footnote{For cool
stars with spectral type B ar later LTE modeling may be adequate but there are always Non-LTE
effects in any star, which are important at least if high-resolution and/or high/energy observations 
are analyzed.}. Such models are available since \citet{werner86,werner89} presented the first calculation
based on the accelerated lambda iteration (ALI) techniques. For more details see, e.g.,
\citet{h03,lh03,rd03}. Presently, our T\"ubingen NLTE Model Atmosphere Package\footnote{
http://astro.uni-tuebingen.de/\raisebox{.3em}{{\tiny $\sim$}}rauch/TMAP/TMAP.html\hfill\hbox{}}
\citep[\emph{TMAP},][]{werner03,rd03} is capable to calculate plane-parallel and spherical,
chemically homogeneous, Non-LTE model atmospheres in radiative and hydrostatic equilibrium and 
considers opacities of all species from hydrogen to nickel \citep{rauch97,rauch03}.

In the last two decades, 
\emph{TMAP} has been successfully employed for the analysis of hot post-AGB stars \citep[e.g\@.][]{jahn07,rauch07}.
Space-based observatories like, e.g., FUSE\footnote{Far Ultraviolet Spectroscopic Explorer\hfill\hbox{}} and 
HST\footnote{Hubble Space Telescope\hfill\hbox{}}, have provided high-resolution, high-S/N UV spectra which
have shown that we arrived at a severe limitation due to the lack of reliable atomic and line-broadening data
for highly ionized species. This is a challenge for atomic physics.

\section{Selected results of spectral analyses}
\label{sect:analysis}

In this section, we will highlight some recent important results of our analyses. For a more detailed review on
the spectroscopy of hydrogen-deficient post-AGB stars see \citet{werner06}.
\vspace{3mm}

\noindent
{\bf Fluorine} (F\,{\sc vi} $\lambda\,1139.50\,\mathrm{\AA}$) has been identified for the first time in 
FUSE observations of post-AGB stars \citep{wrk2005}. 
F is produced during helium burning on the AGB via 
$^{14}\mathrm{N}(\alpha,\gamma)^{18}\mathrm{F}(\beta^+)^{18}\mathrm{O}(\mathrm{p},\alpha)^{15}\mathrm{N}(\alpha,\gamma)^{19}\mathrm{F}$.
Our spectral analysis has shown that the F abundance in hydrogen-deficient PG\,1159-type stars is about $200\times$ solar
\citep[cf\@.][]{asplund05}. 
This result confirms the F intershell abundances predicted by evolutionary models \citep{lugaro04}. 
\vspace{3mm}

\noindent
{\bf Neon} (Ne\,{\sc vii} $\lambda\,3644.6\,\mathrm{\AA}$) was first identified by \citet{wr94} in
optical observations of PG\,1159 stars. It is worthwhile to note that this line was
detected in a high-temperature discharge plasma as a weak blend on a strong
Ne\,{\sc ii} line already many years ago \citep{ck99} und is frequently
used as a calibration line in the laboratory \citep{kkk93} -- unfortunately this was not known in astronomy 
before and demonstrates the need for improvement and extension of atomic-data databases.
Ne is synthesized in the helium-burning shell by the
$^{14}\mathrm{N}(\alpha,\gamma)^{18}\mathrm{F}(\beta^+)^{18}\mathrm{O}(\alpha,\gamma)^{22}\mathrm{Ne}$ chain.
The determined photospheric abundance is 2\% by mass \citep[$11\times$ solar,][]{wr94}, well in agreement 
with early evolutionary models of \citet{it85}. 

\begin{figure}
 \includegraphics{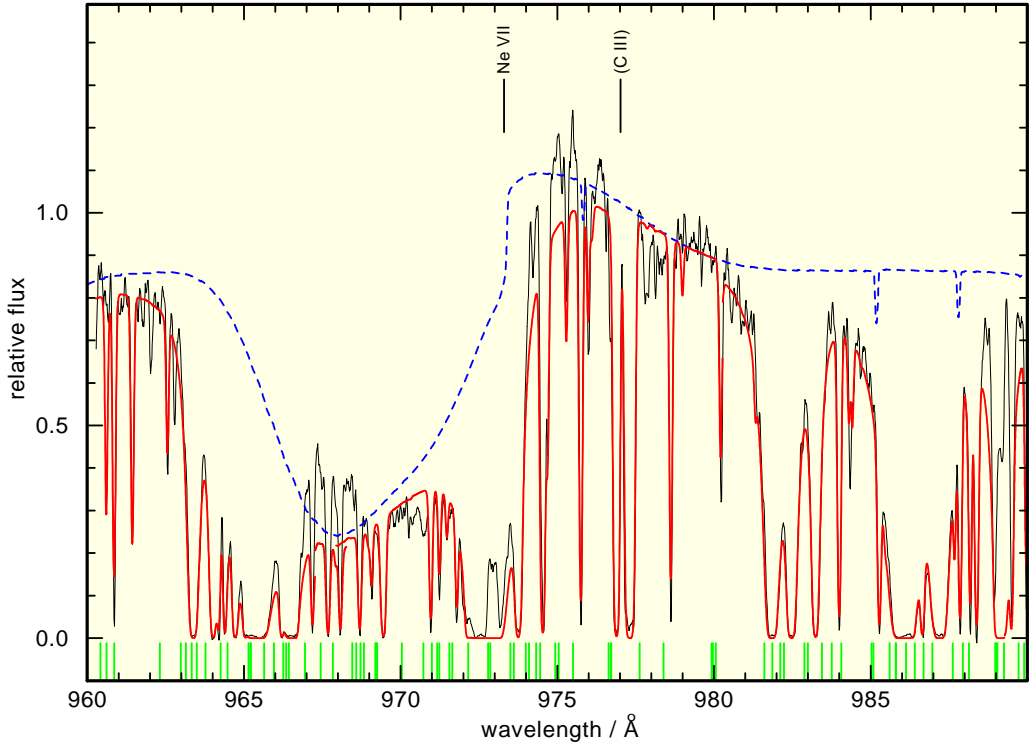}\centering
  \caption{Comparison of the synthetic stellar spectrum (dashed, calculated with the
           \emph{HotBlast} wind code) and the combined synthetic
           stellar and interstellar line spectrum around Ne\,{\sc vii} $\lambda\,973.3\,\mathrm{\AA}$ 
           with the FUSE observation of the CSPN  NGC\,7094. Note that C\,{\sc iii} $\lambda\,977.3\,\mathrm{\AA}$
           is much too weak at $T_\mathrm{eff} = 110\,000\,\mathrm{K}$ to explain the observed (Ne\,{\sc vii}) profile. 
           The vertical bars at the bottom mark the positions of H$_2$ lines in the interstellar spectrum.}
  \label{fig:wind}
\end{figure}

In FUSE observations of PG\,1159 stars, we could firstly identify
Ne\,{\sc vii} $\lambda\,973.3\,\mathrm{\AA}$ \citep{wrk2004}
-- a stellar wind can form a strong Ne\,{\sc vii} P\,Cygni profile 
\citep{hbh05} -- the closely located C\,{\sc iii} $\lambda\,977.3\,\mathrm{\AA}$ 
line is much too weak at the relevant temperature regime (Fig.\,\ref{fig:wind}). 

Recently, \citet{kb06} presented reliable atomic data
of Ne\,{\sc viii}. This enables us to calculate the Ne\,{\sc viii} spectrum of hot post-AGB stars and
we could identify Ne\,{\sc viii} absorption lines in FUSE observations \citep{wrk2007b}.

The identification of Ne\,{\sc vii} and Ne\,{\sc viii} lines provides a new sensitive tool for very hot stars to 
determine $T_\mathrm{eff}$ by an evaluation of the Ne\,{\sc vii}/{\sc viii} ionization equilibrium.
It is worthwhile to note, that $T_\mathrm{eff}$ of the hottest known helium-rich DO white dwarf,
KPD\,0005+5106, had to be revised to $T_\mathrm{eff}\approx 200\,000\,\mathrm{K}$ (previously $120\,000\,\mathrm{K}$) 
by the identification and modelling of Ne\,{\sc viii} emission lines in its spectrum. These lines were previously
thought to be O\,{\sc viii} lines which however could not be of photospheric origin \citep{wrk2007b}.
\vspace{3mm}

\noindent
{\bf Argon} also provides an ionization equilibrium (Ar\,{\sc vi}/{\sc vii}) for the determination of
$T_\mathrm{eff}$ in very hot stars. \citet{wrk2007a} identified  Ar\,{\sc vii} $\lambda\,1063.55\,\mathrm{\AA}$
in FUSE observations of hot central stars of planetary nebulae (CSPNe) and in white dwarfs.
\citet{rauch07} identified Ar\,{\sc vi} lines in a HST/STIS\footnote{Space Telescope Imaging Spectrograph\hfill\hbox{}} 
observations of the CSPN LSV\,46$^\mathrm{o}$21.
\vspace{3mm}

\noindent
{\bf Iron} is not affected by nuclear burning but its abundance may be reduced due to n-captures in
the s-process. Evolutionary calculations predict only a very weak extent of
its depletion. In contrast, spectral analyses of FUSE observations with sufficiently high S/N of four hydrogen-deficient 
post-AGB stars have shown that no iron lines are detectable. Thus, a very strong Fe depletion (at least
$1 - 2\,\mathrm{dex}$) takes place in the intershell \citep{MEA02}.

\section{One example of spectral analysis: \\
         \hspace{6mm}The exciting star of the planetary nebula NGC\,7094}
\label{sect:examples}

The CSPN NGC\,7094 is a so-called hybrid PG\,1159 star, i.e\@. it
exhibits hydrogen lines in its spectrum. In a Non-LTE spectral analysis \citet{dreizler95} determined
$T_\mathrm{eff} = 110\,000\,\mathrm{K}$, $\log g = 5.7$, and 
H:He:C:N:O = 0.42:0.51:0.05:($<$\,0.01):($<$\,0.01) in mass fractions.
The CSPN NGC\,7094 may thus be an AFTP star. \citet{MEA02} discovered a strong iron deficiency of about
two dex. In an on-going analysis of FUSE and HST/STIS observations (Ziegler et al\@. in prep.), we aim 
to identify nickel lines in order to determine the Fe/Ni abundance ratio which is a indicator for the
efficiency of the s-process. We did not succeed in this attempt. It is possible that the s-process
has transferred iron into nickel and then into more heavy species. Unfortunately, we cannot search
for lines of trans iron-group elements because no atomic data are available.

We used the newly developed \emph{HotBlast} 
Non-LTE code for spherically expanding atmospheres in order to calculate the P\,Cygni profile of 
Ne\,{\sc vii} $\lambda\,973.3\,\mathrm{\AA}$. \emph{HotBlast} 
uses as an input the atmospheric structure of our static \emph{TMAP} model to simulate the
atmosphere below the ``wind region''.
In the course of our analysis, it turned out that the FUSE observation is strongly contaminated by
interstellar line absorption (Fig.\,\ref{fig:wind}). Therefore, we employed the programme \emph{OWENS}
\citep[cf\@.][]{Lemoine02, Hebrard02} which can simulate interstellar clouds with
individual parameters like, e.g., radial velocity, column density in the line of sight, temperature of the
gas, and microturbulence velocity. Although we show in Fig.\,\ref{fig:wind} only a qualitative fit of the 
ISM line absorption, it is obvious that simultaneous modeling of both, the stellar and interstellar line
spectrum, is necessary in order to search and identify weak metal lines.

\section{Spectral analysis in the 21$^\mathrm{st}$ century}
\label{sect:gavo}

Spectral analysis by means of Non-LTE model-atmosphere techniques has for a long time been regarded as a
domain of specialists. Within the \emph{German Astrophysical Virtual Observatory} 
(\emph{GAVO}\footnote{http://www.g-vo.org\hfill\hbox{}}) project,
we have created the VO service \emph{TheoSSA}\footnote{Theoretical Simple Spectra Access,
http://vo.ari.uni-heidelberg.de/ssatr-0.01/TrSpectra.jsp}. A VO user may use pre-calculated grids
of spectral energy distributions (SEDs,
in a pilot phase calculated by \emph{TMAP} for hot, compact stars only) which are ready to use
and it may be interpolated between them to match the user-required parameters. This is the easiest way to use
synthetic SEDs calculated from Non-LTE model atmospheres. They represent stars much better than still
oftenly used blackbody flux distributions for PNe analyses. 

If individual parameters are requested which do not fit to an already existing SED in the database, the VO user is guided to
\emph{TMAW}\footnote{http://astro.uni-tuebingen.de/\raisebox{.3em}{{\tiny $\sim$}}rauch/TMAW/TMAW.shtml\hfill\hbox{}}.
With this WWW interface, the VO user may calculate an individual model atmosphere, requesting
$T_\mathrm{eff}$, 
$\log g$, 
and mass fractions $\left\{X_\mathrm{i}\right\}$, $\mathrm{i} \in [\mathrm{H,\,He,\,C,\,N,\,O}]$
(more species will be included in the future).
For this calculation, standard model atoms are used which are provided within the
T\"ubingen Model-Atom Database (\emph{TMAD}\footnote{
http://astro.uni-tuebingen.de/\raisebox{.3em}{{\tiny $\sim$}}rauch/TMAD/TMAD.html\hfill\hbox{}}).
Since the VO user can do this without detailed knowledge of the programme code working in the background,
the access to individually calculated SEDs is as simple as the use of pre-calculated SEDs -- however,
the calculation needs some time (depending on the number of species considered, the wall-clock time is ranging
from hours to a few days). Standard SEDs of all calculated model atmospheres are automatically ingested into
the \emph{GAVO} data base and, thus, it is growing in time.

In case that a detailed spectral analysis is performed, an experienced VO user may create an own atomic
data file tailored for a specific purpose considering all necessary species ($\mathrm{i} \in [\mathrm{H - Ni}]$) 
and calculate own model atmospheres and SEDs.

In close collaboration of \hbox{}\hspace{1mm}\emph{GAVO}\hspace{1mm}\hbox{} with the 
\hbox{}\hspace{1mm}\emph{German Astronomy Community Grid} 
(\emph{AstroGrid-D}\footnote{http://www.gac-grid.de\hfill\hbox{}}), the
calculations of \emph{TMAW} will be performed on GRID computers in the future. 
This will allow to calculate small grids of model atmospheres and SEDs on a reasonable timescale.

\newpage

\begin{acknowledgments}
T\@. R\@. is supported by the German Astrophysical Virtual Observatory (GAVO) project
of the German Federal Ministry of Education and Research (BMBF) under the grant 05\,AC6VTB. 
This work has been done using the profile fitting procedure \emph{OWENS} developed by M\@. Lemoine 
and the FUSE French Team.
\end{acknowledgments}


\begin{thebibliography}{}

\bibitem[Asplund et al\@.(2005)Asplund, Grevesse \& Sauval]{asplund05}
         Asplund, M., Grevesse, N., \& Sauval, A\@. J\@.
         2005,
         in: {\it Cosmic Abundances Records of Stellar Evolution and Nucleosynthesis},
         eds\@. T\@. G\@. III\@. Barnes, \& F\@. N\@. Bash,
         The ASP Conference Series Vol\@. 336 (San Francisco ASP), p\@. 25                                                      

\bibitem[Dreizler et al\@.(1995)Dreizler, Werner, \& Heber]{dreizler95}
         Dreizler, S., Werner, K., \& Heber, U\@.
         1995,
         in: {\it White Dwarfs},
         eds\@. D\@. Koester, \& K\@. Werner,  
         LNP, 443, 160

\bibitem[H{\'e}brard et al\@.(2002)H{\'e}brard et al\@.]{Hebrard02} 
         H{\'e}brard, G., et al\@. 
         2002, 
         ApJS, 140, 103 

\bibitem[Herald et al\@.(2005)Herald, Bianchi \& Hillier]{hbh05}
         Herald, J\@. E., Bianchi, L., \& Hillier, D\@. J\@. 
         2005,
         ApJ, 627, 424 

\bibitem[Hubeny(2003)Hubeny]{h03}
         Hubeny, I\@.
         2003,
         in: {\it Stellar Atmosphere Modeling},
         eds\@. I\@. Hubeny, D\@. Mihalas, \& K\@. Werner, 
         The ASP Conference Series Vol\@. 288 (San Francisco ASP), p\@. 51

\bibitem[Iben \& Tutukov(1985)Iben \& Tutukov]{it85}
         Iben, I., Jr., \& Tutukov, A\@. V\@.
         1985,
         ApJS, 58, 661

\bibitem[Iben et al\@.(1983)Iben et al\@.]{iben83}
         Iben, I., Jr., Kaler, J\@. B., Truran, J\@. W., \& Renzini, A\@.
         1983,
         ApJ, 264, 605

\bibitem[Jahn et al\@.(2007)Jahn et al\@.]{jahn07}
         Jahn, D., Rauch, T., Reiff, E., Werner, K., Kruk, J\@. W., Dreizler, S., \& Herwig, F\@.
         2007,
         A\&A, 462, 281                    

\bibitem[Johnston \& Kunze(1969)Johnston \& Kunze]{jk69}
         Johnston, W\@. D., \& Kunze, H.-J\@. 
         1969, 
         ApJ, 157, 1469

\bibitem[K\"onig et al\@.(1993)K\"onig, Kolk \& Kunze]{kkk93}
         K\"onig, R., Kolk, K.-H., \& Kunze, H.-J\@. 
         1993, 
         Physica Scripta 48, 9

\bibitem[Kramida \& Buchet-Poulizac(2006)Kramida \& Buchet-Poulizac]{kb06} 
         Kramida, A\@. E., \& Buchet-Poulizac, M.-C\@. 
         2006, 
         Eur\@. Phys\@. J\@. D, 39, 173

\bibitem[Lanz \& Hubeny(2003)Lanz \& Hubeny]{lh03}
         Lanz, T., \& Hubeny, I\@.
         2003,
         in: {\it Stellar Atmosphere Modeling},
         eds\@. I\@. Hubeny, D\@. Mihalas, \& K\@. Werner, 
         The ASP Conference Series Vol\@. 288 (San Francisco ASP), p\@. 117

\bibitem[Lemoine et al\@.(2002)Lemoine et al\@.]{Lemoine02} 
         Lemoine, M., et al\@. 
         2002, 
         ApJS, 140, 67 

\bibitem[Lugaro et al\@.(2004)Lugaro et al\@.]{lugaro04} 
         Lugaro, M., Ugalde, C., \& Karakas, A\@. I., et al\@.
         2004, 
         ApJ, 615, L934

\bibitem[Miksa et al\@.(2002)Miksa et al\@.]{MEA02}
         Miksa, S., Deetjen, J\@. L., Dreizler, S., Kruk, J\@ .W., Rauch, T., \& Werner, K\@. 
         2002,
         A\&A 389, 953 

\bibitem[Rauch(1997)Rauch]{rauch97}
         Rauch, T\@
         1997,
         A\&A, 320, 237
     
\bibitem[Rauch(2003)Rauch]{rauch03}
         Rauch, T\@
         2003,
         A\&A, 403, 709
     
\bibitem[Rauch \& Deetjen(2003)Rauch\& Deetjen]{rd03}
         Rauch, T., \& Deetjen, J\@. L\@.
         2003,
         in: {\it Stellar Atmosphere Modeling},
         eds\@. I\@. Hubeny, D\@. Mihalas, \& K\@. Werner, 
         The ASP Conference Series Vol\@. 288 (San Francisco ASP), p\@. 103

\bibitem[Rauch et al\@.(2007)Rauch et al\@.]{rauch07}
         Rauch, T., Ziegler, M., Werner, K., et al\@.
         2007,
         A\&A, 470, 317 

\bibitem[Werner(1986)Werner]{werner86}
         Werner, K\@
         1986,
         A\&A, 161, 177
     
\bibitem[Werner(1989)Werner]{werner89}
         Werner, K\@
         1989,
         A\&A, 226, 265
     
\bibitem[Werner \& Herwig(2006)Werner \& Herwig]{werner06}
         Werner, K., \& Herwig, F\@.
         2006,
         PASP, 118, 183     

\bibitem[Werner et al\@.(2003)Werner et al\@.]{werner03}
         Werner, K., Deetjen, J\@. L., Dreizler, et al\@.
         2003,
         in: {\it Stellar Atmosphere Modeling},
         eds\@. I\@. Hubeny, D\@. Mihalas, \& K\@. Werner, 
         The ASP Conference Series Vol\@. 288 (San Francisco ASP), p\@. 31

\bibitem[Werner \& Rauch(1994)Werner \& Rauch]{wr94}
         Werner, K., \& Rauch, T\@.
         1994,
         A\&A, 284, L5

\bibitem[Werner et al\@.(2004)Werner et al\@.]{wrk2004}
         Werner, K., Rauch, T., Reiff, E., Kruk, J\@. W., \& Napiwotzki, R\@.
         2004,
         A\&A, 427, 685

\bibitem[Werner et al\@.(2005)Werner, Rauch \& Kruk]{wrk2005}
         Werner, K., Rauch, T., \& Kruk, J\@. W\@.
         2005,
         A\&A, 433, 641

\bibitem[Werner et al\@.(2007a)Werner, Rauch \& Kruk]{wrk2007a}
         Werner, K., Rauch, T., \& Kruk, J\@. W\@.
         2007a,
         A\&A, 466, 317

\bibitem[Werner et al\@.(2007b)Werner, Rauch \& Kruk]{wrk2007b}
         Werner, K., Rauch, T., \& Kruk, J\@. W\@.
         2007b,
         A\&A, 481, 807

\end{thebibliography}
\end{document}